# Leveraging user profile attributes for improving pedagogical accuracy of learning pathways

Tanmay Sinha
Department of Computer Science
Vellore Institute of Technology
Chennai 600127, India
tanmay.sinha655@gmail.com

Ankit Banka
Department of Computer Science
Vellore Institute of Technology
Chennai 600127, India
ankitbanka17@gmail.com

Dae Ki Kang[*]
Department of Computer Engineering
Dongseo University
Busan 617716, Korea
dkkang@dongseo.ac.kr

*Abstract* - In recent years, with the enormous explosion of web based learning resources, personalization has become a critical factor for the success of services that wish to leverage the power of Web 2.0. However, the relevance, significance and impact of tailored content delivery in the learning domain is still questionable. Apart from considering only interaction based features like ratings and inferring learner preferences from them, if these services were to incorporate innate user profile attributes which affect learning activities, the quality of recommendations produced could be vastly improved. Recognizing the crucial role of effective guidance in informal educational settings, we provide a principled way of utilizing multiple sources of information from the user profile itself for the recommendation task. We explore factors that affect the choice of learning resources and explain in what way are they helpful to improve the pedagogical accuracy of learning objects recommended. Through a systematical application of machine learning techniques, we further provide a technological solution to convert these indirectly mapped learner specific attributes into a direct mapping with the learning resources. This mapping has a distinct advantage of tagging learning resources to make their metadata more informative. The results of our empirical study depict the similarity of nominal learning attributes with respect to each other. We further succeed in capturing the learner subset, whose preferences are most likely to be an indication of learning resource usage. Our novel system filters learner profile attributes to discover a tag that links them with learning resources.

*Keywords - Apriori principle, Clustering, NMF, Learning Pathway, Pedagogical accuracy, Technology enhanced learning.*

## I. Introduction

Education has been acknowledged as one of the major pillars to support and improve lives of the economically disadvantaged [26]. Technology enhanced learning(TEL) aims to enhance learning practices of individuals and organizations by the design, development and testing of interactive socio technical innovations [2]. It provides an opportunity to unleash the full potential of machine learning algorithms and apply them to quantify and incorporate psychological factors that affect learning. As the web is becoming more ubiquitous, an increased number of learners are gradually turning towards online education to seek additional support and guidance. Therefore, improving learner experience is also another major focus of TEL. Over the years, various learning repositories like Khan Academy, Merlot(with more than 20,000 learning resources and about 70,000 registered users), OER Commons(with about 18,000 resources), European Schoolnet's Learning Resource Exchange(with more than 43,000 learning resources from 25 different content providers in Europe and beyond), online education platforms like Coursera, Udacity, EdX, Khan Academy, Venture Lab and e-learning platforms NPTEL have been established. They provide personalized learning environments and offer pedagogical support for learners in informal educational settings.

Apart from providing custom self paced learning tools, such organizations also measure student progress and provide targeted interventions. In contrast to formal education systems in schools and universities where predefined course structures, well formed accreditation procedures and a regular quality check of knowledge inflow and outflow is maintained by domain experts, informal learning is basically a kind of lifelong learning activity. Here, students are responsible for their own learning pace and path [1]. The process of learning is often self directed and heavily dependent on users preferences. Because of no directed course plan and massive enrollment to form such informal learning networks, these learners are often overwhelmed with a plethora of alternative learning resources to choose from. It is highly impossible for a faculty to cater to the individual problems of learners. As pointed out in [25], the problem of improving the quality of education is quite multifaceted and intricate. However, the suggestion of correct pathways to be followed while studying important concepts, can benefit these learners greatly. Though the field of TEL Recommender system(TEL RecSys) [2] addresses this scenario of mass customization, the major factor for deciding the most appropriate learning activities is still learner's rating, review of resources and his fundamental demographic information. This may work well for e-commerce recommender systems, but similar applications in the domain of learning may not be suitable.

Basically there are the two major differences between product recommendations and TEL RecSys that need to be considered. Firstly, learning is an activity that takes more time, considering the different study goals of individuals and their grasping capabilities. Also, the end objective may be variable.

---

[*] Corresponding author: Tel.: +82513201724, Email: dkkang@dongseo.ac.kr

Learners may wish to have different accomplishments after completion of a learning resource. Secondly, the most preferred(rated) learning path may not be pedagogically the most adequate [3]. Thus, the educational specific needs and information seeking tasks of stakeholders, make the design of learning resource recommendation quite complex [16].

Our statement of contribution in this paper is to develop a system that can effectively categorize resources based on inherent factors that affect student's learning. We aim to extract such learning domain specific attributes of learners, that have been discovered from past theoretical research. Such critical factors belong to an entirely different context and play a much more significant role, as compared to basic profile information like name, age, sex and other demographic factors. We believe that our system has the potential of improving alignment of learner's preferences with the learning resources they are using. Also, it could increase learning efficiency, through a more structured guidance that is well customized and adapted to the learner's objectives.

This paper is structured as follows: In section II.I, the technical terms used are defined. Then, in the following sections II.II and II.III, the user profile attributes are described and challenges are presented. The hypothesis, problem statement and the proposed method is described in detail in section III. Section IV talks about the experimental results and discussions. Prior work is discussed in section V. The conclusion and future work follow at the end in section VI and VII.

## II.I. PRELIMINARIES

1)**Learning Management system(LMS)-** It is a system that assists learners in finding learning resources and guides them to make the best choices in managing these resources. It also supports these learners in achieving their goals in a specific domain. A good literature review of LMS can be found in [4].
2)**Learning pathway-** A set of learning objects like books, multimedia(audio recordings, videos), images, slides etc which are packaged and organized in a sequence form a learning pathway. A formal definition of the same can be found in [5].
3)**Transfer Learning-** The process of applying concepts and skills learned in one domain to another related or disjoint domain in different scenarios is called transfer learning [6]. The absence of a guide to assist learners in providing a good cross domain application context, makes an understanding of the research issues associated with transfer learning quite difficult.
4)**Pedagogical adequacy(accuracy)**- It is the kind of education that focuses equally on five different aspects of learning- student centred tutelage with an understanding of his educational background, strategic course instruction design to maximize learning, evaluation of student's advancement towards the course learning objectives for knowledge retention, use of different teaching strategies, awareness regarding teaching challenges and knowing ways to handle them[7].
5)**K-means clustering**- Intuitively, what clustering techniques aim to achieve is finding data points which are more similar to one another than the rest of the points in n-dimensional feature space. K-means is a widely adopted clustering method that starts out by creating 'k' different clusters and assigning each data point to one of the clusters based on some standard distance measure like Euclidean, Pearson correlation etc. At every iteration, centroid of the clusters are updated with the mean value of the data points assigned to that cluster. The idea is to minimize intra cluster distance and maximize inter cluster distance. Proposed in [22], K means clustering has been a widely studied unsupervised machine learning algorithm.
6)**Apriori principle**- Originally presented in [23], it optimizes association analysis in large scale machine learning. Either for finding out frequent itemsets or discovering association rules in large datasets, apriori principle is applied to reduce the number of possible interesting itemsets. Measures like support and confidence levels are established to evaluate the associations discovered. Apriori principle says that if an itemset is infrequent, all it's supersets will also be infrequent. At every iteration of the algorithm, candidate itemsets are pruned if their support levels are below the minimum support. The process of joining lower order itemsets based on this criteria is repeated, until no more itemsets can be created further.
7)**Non negative matrix factorization(NMF)**- Formally defined in [24] for learning parts of objects and combining them to get an aggregate representation, NMF derives it's roots from linear algebra. The basic idea of NMF is that it decomposes matrix A into two matrices B(features) and C(weights) and also enforces a non negative constraint on the entries of these matrices. In machine learning, it is used as a feature extraction algorithm, where the decomposed matrices represent the extracted features in the observations and their importance or relevance to the observed data.

## II.II. USER PROFILE ATTRIBUTES

User model is a distinctive feature of adaptive systems, as this information allows the system to behave differently for different users [20]. In our work, we consider the following user profile attributes for incorporating into the proposed machine learning framework to tag learning resources. The objective is to improve the pedagogical accuracy and adequacy of TEL RecSys, by suggesting resources that reflect the learner's objective and form a close match to his learning requirements. By doing so, the learner can be kept motivated to complete his learning activities in an effective and efficient manner.

1)**Skill level-** This includes current competency level of the learner and the target skill level that is to be achieved. Based on Bloom's taxonomy and CDIO report[8], the following proficiency levels are identified:

- Level 1- To have experienced or been exposed to(Knowledge)- refers to the process of recalling primer knowledge on a subject
- Level 2- To be able to participate in and contribute to(Comprehension)- refers to the ability to describe and define what is learnt
- Level 3- To be skilled in the practice or implementation of(Application)- refers to the process of execution of concepts
- Level 4- To be able to understand and explain(Analysis)- refers to the process of analysis, interpretation and discussion on results
- Level 5- To be able to lead or innovate in(Synthesis)- refers to the process of creation of thoughts and development of new ideas
- Level 6- To be able to assess and evaluate(Evaluate)- refers to the process of estimation and evaluation of the concepts learnt

2)**Preferred Learning strategy-** This basically refers to the intended way of understanding information, solving problems, checking and evaluating the knowledge gained. Based on the strategic instruction model proposed by [9], we have included the following five learning strategies for the purpose of our study:
- Assignment Completion Strategy- designed to enable students to complete and hand in the assignments on time
- Test-Taking Strategy- designed to be used while taking classroom(physical/virtual) tests, with students allocating time and certain priority to each section of the test
- Mnemonic Strategy- designed to help students remember complex concepts using simple representations(diagrams) and mnemonic codes.
- Self-Questioning Strategy- designed to help students create their own motivation while reading by creating questions in mind, predicting answers, searching, verifying and paraphrasing them
- Error Monitoring Strategy- designed to help students independently detect and correct errors
- Teamwork Strategy- designed to provide a framework for organizing and completing specific tasks in small groups

3) **Available Learning time -** This includes the number of time units(hours per day and number of weeks), and indicates the time allotted for the purpose of study in an informal educational setting.

4)**Preferred Presentation style-** Because of the difference in cognitive processing, we consider visual, textual and auditory presentation styles. More specifically, we include:
- Portable document format(pdf) documents
- Power points
- Web pages(html)
- Videos
- Audio books and resources

An alternative to the above styles include presentation of examples, presentation of theoretical knowledge and practical examples [10]. Over and above all these explicit factors that can be gathered by an initial questionnaire, other implicit factors like effort of the learner and his willingness to obey and follow the recommendations can be measured during course progress [11][12].

## II.III. CHALLENGES AND MOTIVATION

Though technology alone cannot solve complex problems like providing universal and high quality education, but it can surely improve the quality of solutions and their impact on learning. This is the driving factor behind our work.

1)We notice that a major issue in TEL RecSys is the use of only learner ratings, reviews and basic profile information for providing tailored learning pathways. Other critical factors related to student's academic profile and interests are left out. This leads to inaccurate and unsatisfactory recommendations most of the times. However, because learning takes place in extremely diverse and rich environments, we include the other dimensions as stated in Section II.I to make the learning model more helpful and close to real life behaviour. Because these user specific attributes do not have a direct link with the learning resources and due to the unavailability of tagged data for these resources, there is a need to incorporate such parameters into the recommendation framework. And our proposed solution finds a way to fill this missing link by using machine learning algorithms to accomplish the task at hand.

2)The other challenge while applying machine learning to make appropriate sense of the user profile attributes is that, though skill levels and learning time can be inferred numerically, learning strategy and presentation style are very nominal in nature. Deriving similarities between them is quite difficult numerically. We cannot represent such nominal variables on a continuous scale and perform arithmetic operations on them. For example, it's difficult to comprehend, how similar or dissimilar test taking and mnemonic learning strategies are, without any training data. This fact has useful implications, because distance measures like Euclidean,

Pearson correlation etc. can't be applied to address such a situation. Thus, there is a need to characterize the data further to leverage these nominal attributes.

## III. METHODOLOGY

Our methodology for tagging learning resources consists of deriving a subset of learners who have rated the learning resource above a particular threshold, finding a relation between their nominal attribute parameters, grouping these learners to find the cluster of largest size, and using the most frequently occurring learning profile attributes as a more informative tag for the learning resource.

### III.I. HYPOTHESIS

In order to arrive at the formal algorithm design and answer the compound key question of our research, i.e., how to establish the link between learning resources and user profile attributes that play a major role in learning(as discussed above), we formulate our research hypothesis as follows :
**H1**: Learning resources, apart from being distinguishable solely by learner ratings, can be more efficiently categorized and tagged according to learner study motives.
**H2**: Incorporating user centric attributes for suggesting customized learning pathways leads to pedagogically more accurate and satisfactory recommendations for the learner.

We address the first hypothesis in this paper and formalize it as follows:
Let 'S' be a System
$L=\{l_1, l_2, l_3, \ldots, l_m\}$ be a set of learning resources.
$U=\{u_1, u_2, u_3, \ldots, u_n\}$ be a set of learners.
$A=\{a_1, a_2, a_3, a_4, a_5\}$ be a set of learner profile attributes.
$R=\{1,2,3,4,5,\ldots,10\}$ be set of possible ratings.
where
$m, n \in N$
$a_1 \rightarrow$ current skill level $\in \{1,2,3,4,5,6\}$
$a_2 \rightarrow$ target skill level $\in \{1,2,3,4,5,6\}$, $\forall u_i \in U$, $a_2 > a_1$
 (i=1 to n)
$a_3 \rightarrow$ learning strategy $\in \{1,2,3,4,5\}$
  /*nominal attribute*/
$a_4 \rightarrow$ preferred presentation style $\in \{1,2,3,4,5\}$
  /*nominal attribute*/
$a_5 \rightarrow$ learning time $\in N$   /*discretized as different ranges*/
$p_{ij} \rightarrow$ user $u_i$'s profile attribute $a_j$ (i=1 to n, j=1 to 5)
$r_{iy} \rightarrow$ user $u_i$'s rating of learning resource $l_y$, (i=1 to n, y=1 to m)
**Problem Statement:**
$\forall l_y \in L$, tag $l_y$ with each $a_j \in A$, (y=1 to m, j=1 to 5)

### III.II. SOLUTION & APPROACH FOLLOWED

**Step 1- Alg. *"build_learner_subset"***
*input- learner set*
*output- learner subset based on rating threshold criteria*
a. $\forall l_y \in L$ build $V \subseteq U \mid r_{iy} \geq \delta_0$, $\delta_0 \leq 10$ and $\delta_0 \in N$ (i=1 to n, y=1 to m)   /*build learner subset*/
b. $\forall u_i \in V$, consider $p_{ij}$ as a point in 5-D feature space (i= 1 to n, j=1 to 5)

**Fig (i) Alg. *"build_learner_subset"***

In Fig (i), for all learning resources, we get a list of learners who have rated them greater than a threshold value (e.g.- greater than 9/10). This list is an indicative of learners who have preferred those resources over and above others. For experimentation, we tune this threshold value, so that we get a learner subset that is big enough to be tested for future steps.

**Step 2- Alg. *"nominal_attributes_quantify"***
*input- number of learners using nominal attribute parameter pair (p,q) in common*
*output- numerical values for nominal attribute parameters*
$\forall$ nominal attributes($a_3$, $a_4$)
 a. Construct matrix $A_{5*5}$ with each row and column representing specific nominal attribute parameters, and $A_{ij}$ = no. of users who have parameter i and j in common.
 b. Apply NMF to split A into B & C:
 $A_{5*5} = B_{5*k} \times C_{k*5}$
 where $C_{k*5} \rightarrow$ features matrix (rows= features, columns= nominal attribute parameters), and $C_{ij}$= weight of each nominal attribute parameter to the feature extracted,
 and $B_{5*k} \rightarrow$ weight matrix (rows= nominal attribute parameters, columns=features), and $B_{ij}$= weight of each feature in nominal attribute parameter
 c. Now,
 (i) $\forall$ rows in B,
  A. Consider maximum(value).
  B. Choose feature row in C, corresponding to max (value).
  C. Obtain an ordering that defines how similar (important) each nominal attribute parameter is to the other.
  D. Construct matrix D where each row identifies one linear ordering .
 (ii) Make matrix D symmetric by replacing $D_{ij}$ with minimum($D_{ij}$, $D_{ji}$).
 (iii) $\forall$ rows in D, take average score
  (Each average value represents possible numerical values that the nominal attribute can take).

**Fig (ii) Alg. *"nominal_attributes_quantify"***

Secondly, we apply NMF as shown in Fig (ii). NMF will extract features from nominal attribute parameters and output how much importance does each feature have with respect to the other parameters in terms of weights. And, because nominal attribute parameters cannot be both similar to each other by different numerical metrics, we make matrix D symmetric. The average of each linear ordering in matrix D indicates an average similarity with other nominal attribute parameters. Thus, the output of this step is a numerical value for each nominal attribute parameter, which can be used in the next step.

---

**Step 3.1- Alg. *"group_learner"***
*input- data points for every learner, with each coordinate representing different profile attributes*
*output- cluster with largest number of learners*

---

$S \rightarrow \{\phi\}$
Pick first point $p_{ij}$ at random
$S \rightarrow S \cup \{p_{ij}\}$
n=1
choose a value $k \in N$    /*k=number of clusters*/
while $n \leq k$    /*clusters chosen already*/
{
Add point $q'$ to S, such that
$q' \in p_{ij}$ and min{distance($q'$,S)} > distance($q_i$,S)
($q_i \in p_{ij} - q'$, i=1 to n, j=1 to 5)
$S \rightarrow S \cup \{q'\}$
n=n+1
}

**Step 3.2**
While any point has changed in cluster made:
/*K nearest neighbour based clustering*/
{
  $\forall\ p_{ij}$ (i=1 to n, j=1 to 5)
    $\forall\ S_a \in S$ (a=1 to k)
      d($S_a$, $p_{ij}$) = $\sqrt{(S_{a1} - P_{i1})^2 + \ldots + (S_{a5} - P_{i5})^2}$
      /*Euclidean distance metric*/
    x[a]=min{d($S_a$, $p_{ij}$)}
    get l=a which corresponds to min(x[])
    $S_l = S_l \cup p_{ij}$
  $\forall$ every $S_l$, calculate mean of points (M) in that cluster
    $S_l \leftarrow$ Mean(M)
}

**Step 3.3**
Reduce 'k' till average diameter D increases by a very large factor   /*choose optimal no. of clusters*/
D= max(d(p, q)) $\forall$ (p, q) $\in p_{ij}$

**Step 3.4**
Choose cluster with largest number of attribute itemsets
$C \leftarrow$ max{Size($S_a$)}

**Fig (iii) Alg. *"group_learner"***

---

Now, as shown in Fig (iii), in the subset of learners identified, we apply clustering to group learners based on user profile attributes(discussed in Sec II.I). Each learner is a five dimensional point and the coordinates represent his attributes. We then pick the cluster having largest size. This cluster further refines the users according to certain set of user profile attributes. Intuitively, it is a representative of certain pattern in learning behaviour.

---

**Step 4- Alg. *"frequent_attributes_filter"***
*input- data points of the largest cluster chosen*
*output- most frequently occurring user profile attributes*

---

Choose support level sl $\in \{0,..,1\}$ and sl $\in R$
   /*association analysis*/
$L_1 \leftarrow$ {1 itemsets}
for(k=2, $L_{k-1} \neq \phi$, k++)
{
  $c_k \leftarrow \{c | c = a \cup \{b\} \wedge a \in L_{k-1} \wedge b \in \bigcup_k L_{k-1} \wedge b \notin a\}$
  $\forall\ t_i \in C$   /*for each learner record $t_i$ in the cluster*/
  {
    $c_t \leftarrow \{c | c \in c_k \wedge c \subseteq t_i\}$
    $\forall\ c \in c_t$, c.count++; /*maintain count of frequency*/
  }
$L_k \leftarrow \{c \in c_k | c.count > sl\}$   /*include learner itemsets with count greater than min. support level*/
}
return L= $\bigcup_k L_k$
/*returns most frequently occurring learner profile attributes*/

**Fig (iv) Alg. *"frequent_attributes_filter"***

Next, the focus is to quantify this pattern and address the question of tagging learning resources based on the largest cluster chosen. The solution to this problem is application of association analysis to extract hidden knowledge and interesting relationships from this processed data(Fig(iv)). We use frequent itemsets to find the most commonly attributes that occur together using the apriori principle. A minimum support level is defined as an input to this fourth step. The output is a collection of profile attributes, among which we pick the one with largest frequency. Finally, we assign this most frequently occurring and largest itemset of attributes to the learning resource.

V. EXPERIMENTAL RESULTS AND DISCUSSION

The main contribution of our work is the application of machine learning techniques to develop a novel system that can aid in more accurate learning resource pathways recommendation. The strength of the system lies in assisting new learners by providing more relevant and structured navigational support for informal education settings. Instead of "People similar to you who liked learning resource X also liked Y", recommendations take the form of "People similar to you with learning strategy 3, current skill level 2, target skill level 5, preferred presentation style 4, available learning time [20-30] time units who liked the learning resource X also liked Y". The advantages are:

1) Weight of recommendation and it's specificity is increased.
2) A sense of faith in the recommender system is introduced because peer learners having similar competencies are matched.
3) The learner is placed in a more comfortable and confident position to choose his learning resources.
4) Also, learning resources get organized according to learner characteristics.

For the implementation, we used the "Book Crossing" dataset [14], consisting of 2,78,858 learners providing 1,149,780 ratings about 2,71,379 learning resources(books). Because TEL datasets having additional and explicit information about learning attributes are very rare to find, the implementation phase consisted of generating random values for various user profile attributes, as mentioned in Section II.II, and augmenting them with the "Book Crossing" dataset. Then, on applying NMF and extracting 10 independent features, we achieved the following results for the two nominal attributes:

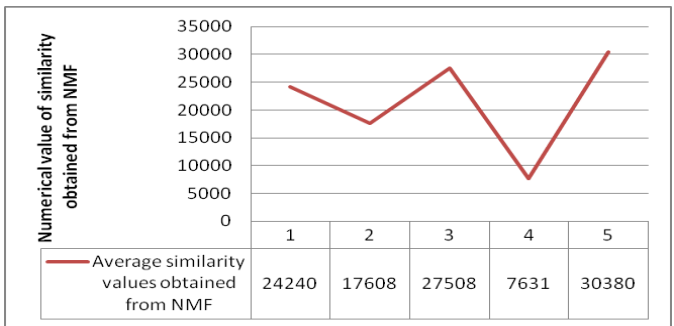

**Fig (v) Learning strategy**

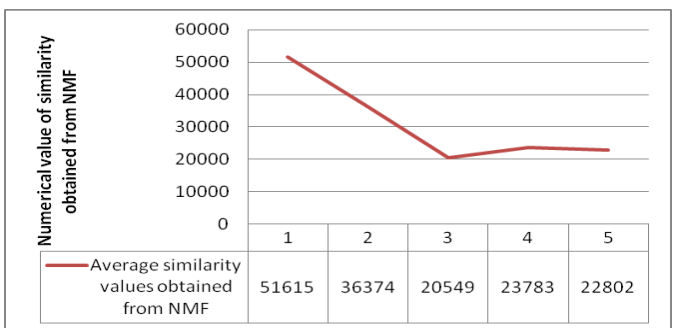

**Fig (vi) Presentation style**

From Fig (v) and Fig (vi), we infer that learning strategy 3 and 5 are most similar or important with respect to each other, while learning strategy 4 and 5 are least similar. Also, among the preferred presentation styles, 4 and 5 are most similar(least distance between them), while 1 and 3 are the least similar(maximum distance).

Some may wonder that it's difficult to evaluate such a system. But, the identified way of evaluating this randomized simulation is by implicit inference of whether the target learner was recommended learning resources based on full user profile cloud of similar learners and in the same domain. The only shortcoming is that we cannot explicitly infer user satisfaction and correctness of the proposed learning pathway through a questionnaire. However, the importance and relevance of implicit inference lies in the fact that it can be collected more easily and at lower cost to the user, although inferences about learning resource desirability will be slightly less accurate than explicitly supplied feedback [15].

Next, we grouped the subset of users, who rated learning resources greater than 6(on a scale of 10). Using intelligent multi attribute visualization techniques, we plotted the clustering results for learning resource with ISBN "000649840X" and ISBN "0684867621" in the dataset, and came up with interesting observations. In Fig (vii), each attribute is represented in a vertical line, where the maximum and minimum values of that dimension are scaled to the upper and lower points on these vertical lines. For 5 visualized attributes, 4 lines connected to each vertical line at the appropriate dimensional value represent a 5-dimensional point. The plot shows few prominent horizontal lines(data points), which intuitively represent strong candidates for tagging learning resources. Coming to Fig (viii), it provides further enhancements to the visualization by depicting data points that are attracted to anchors with value dependent positions. The data instances are placed inside a 5-sided polygon because the clustering process is based on 5 different learner profile attributes.

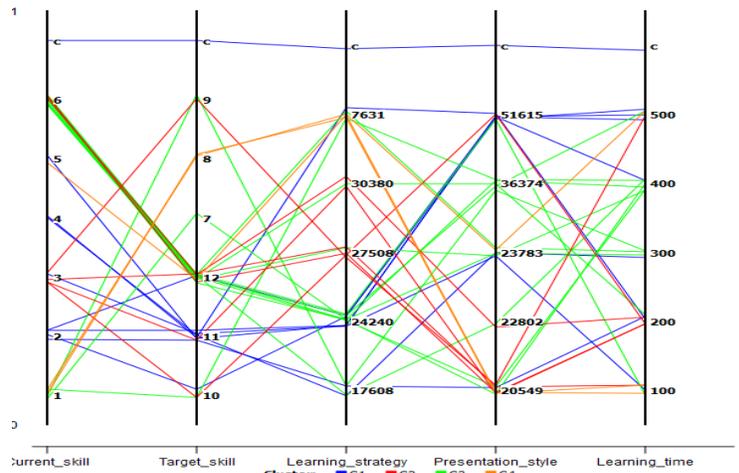

**Fig (vii) Parallel Coordinates Visualization ( for ISBN "000649840X")**

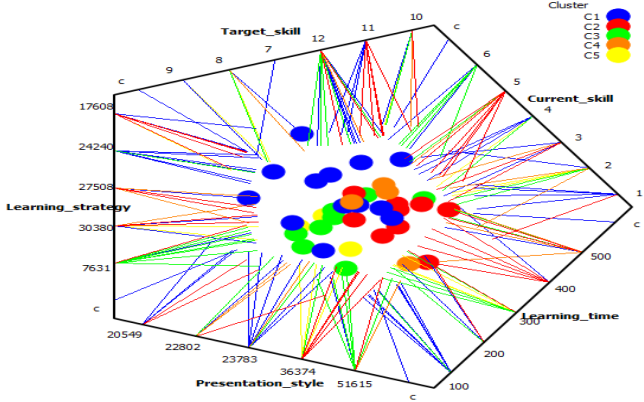

**Fig (viii)Polyviz Visualization(for ISBN "0684867621")**

Next, on applying association analysis to the data points of the largest cluster, we get the certain tags based on most frequently occurring attribute itemsets. Keeping the support level constant at 0.1, we generate tags for each of the learning resources, four of which are shown in Fig (ix). If there are many tags having same frequencies, we generate tag clouds for the corresponding learning resources, where each cloud can hold more than one kind of preference(tag).

Though some of the current tags do not include all 5 user profile attributes, we claim that the system will generate better and higher order tags, once more number of learners start using the system. This is because the input learner subset will grow in size at each step of the system and this will increase the number of possible combinations of learner profile attributes. The justification seems valid, because our proposed system is a bottom up system which incrementally improves as learners and their ratings increase. It is feasible for lifelong learning scenarios with changing learning actions.

| **Learning Resource ISBN** | **Tag-Cloud generated** |
|---|---|
| 000649840X | [6, 6, [41-50], 24240, 20549] |
| 002542730X | [4, 5, -, 30380,23783] <br> and <br> [-, 6, [51-60], 24240, 22802] |
| 038550120X | [6, 6, [51-60], 24240, 23783] <br> and <br> [6, 6, [21-30], 17608, 23783] |
| 0060928336 | [-, -, -, 30380,22802] <br> and <br> [-, 6, [41-50], -, -] |

**Fig(ix)Tags generated for learning resources(5 fields in tag cloud are Current Skill level, Target skill level, Learning time(range in hrs), Learning strategy & Presentation style)**

## V. RELATED WORK

In this section, we describe the prior studies that have been made in the context of development of learner models for educational adaptive hypermedia. In [17], Sicilia et al. carried out an empirical study on the Merlot repository dataset and concluded that ratings cannot correctly differentiate learner preferences. Results showed relatively high mean absolute errors of around 1.0 in a scale of 1 to 5. Psychological studies have stressed on the need to include personalization factors like collective intelligence, competencies of peer learners and learner's reaction towards the learning objects recommended, for inclusion in the design of TEL RecSys [18]. Instead of focusing on development of better algorithms, such learner centric factors can improve accuracy, as learners are both producers of data and recipients of information. A theoretical study on learner models like stereotype, overlay, differential, perturbation and plan models was made in [21], and a generic approach to build these models was provided. However, no algorithms or implementation details were described.

In [19], Verbert et al. did a survey on context aware recommender systems and classified the context framework of TEL RecSys on the basis of computing, location, time, activity, physical conditions, resource and users. The study showed that less than half of the TEL RecSys actually took computing context into consideration. For a majority of them, the implementation was still in the prototype phase. Learner characteristics like basic personal information, prior knowledge level(performance), interests, learning goals, cognitive style and affects(emotions) were identified as key points for consideration in providing personalized learning services. In [12], these determinants were implemented in NetLogo simulation environment, with an objective of identifying the preference and competence gap of learners. Formulas for implicit and explicit learning variables laid the basis of determining the percentage of graduates, their satisfaction levels and time taken to graduate. However, to the best of our knowledge, the identified user profile attributes have not been used for improving pedagogical accuracy of learning pathways. And because this a relatively new area, not much work has been done in applying machine learning oriented approaches to leverage such user tagged parameters. So, it is not trivial to use the previous work along this line to directly serve as subcomponents in our setting.

## VI. CONCLUSION

In this paper, we have designed a solution to cater to learning resource recommendations that inaccurate and less relevant. This disturbs learners and wastes their precious time in searching for the apt learning resources. We call this area of technology enhanced learning sensitive because today, many people from different demographics having different competencies, different ways of learning and interpretation, collaborate on online education portals and leverage resources from learning repositories to improve their skill set. Because of the inefficiency of human support that can be provided to such a large learning network, intelligent ways are required to match learner's actual preferences with learning resources. And leveraging user specific learning attributes is one of the ways to model real life learner behaviour and improve the pedagogical accuracy of suggested learning pathways.

## VII. FUTURE WORK

For future work, we aim to address hypothesis H2, as explained in Section III.I. We also plan to evaluate our system on "SNAP Amazon product co purchasing network" metadata [13], because it is more informative than the current dataset used in experimentation. Furthermore, as the system becomes bigger, accuracy could be tested by dynamically increasing support levels and getting top 'k' patterns of the highest order. The tags achieved at higher support levels would be better than the ones at lower support levels. The goal is to choose an optimal support level, so that order of attribute itemsets is as high as possible, and the number of those itemsets having same frequency is as small as possible. This will reduce ambiguity and control the trade off between which and how many attribute itemsets to use as a tag for the learning resource. Another possible research direction is to extend this simulation based model to develop real world datasets for TEL RecSys. Such an augmented labelling of learning resources will be more reflective of learner proficiencies and objectives.

## ACKNOWLEDGMENT

This work was supported by the National Research Foundation of Korea(NRF) grant funded by the Korea government(MEST) (No. NRF-2013R1A1A2013401).